\documentclass[aps,prd,nofootinbib,showkeys,floatfix,twocolumn]{revtex4-1}
\usepackage{amsmath}
\usepackage{amssymb}
\usepackage{graphicx}
\usepackage{xspace}
\usepackage{bbm} 
\usepackage{color}
\usepackage{colortbl}
\usepackage[utf8]{inputenc}
\usepackage{cancel}
\usepackage{slashed}
\usepackage{soul}
\usepackage{xcolor,colortbl}

\definecolor{mygray}{gray}{0.8}

\definecolor{mkgreen}{rgb}{0.2,.70,.3}
\definecolor{myblue}{cmyk}{0.65, 0.37, 0.0, 0.19}

\newcommand\SPheno{{\tt SPheno 4.0.0}\xspace}

\newcommand\checkmate{{\tt CheckMATE}\xspace}

\newcommand{\lam}{\lambda}
\newcommand{\lsim}{\stackrel{<}{\sim}}
\newcommand{\gsim}{\stackrel{>}{\sim}}

\def\met{\slash\hspace*{-1.5ex}E_{T}}

\usepackage[pdftex,colorlinks,linkcolor=blue,citecolor=blue,urlcolor=blue]{hyperref} 

\definecolor{tobycolour}{rgb}{.6,.0,.4}

\begin{document}
\vspace{1cm}

\title{\Large R-Parity Violation and Direct Stau Pair Production at the LHC}

\hfill \parbox{5cm}{\vspace{ -1cm } \flushright BONN-TH-2020-01  \\DESY 20-007}

\newcommand{\AddrBonn}{%
Bethe Center for Theoretical Physics \& Physikalisches Institut der 
Universit\"at Bonn,\\ Nu{\ss}allee 12, 53115 Bonn, Germany}
\newcommand{\AddrDESY}{%
	DESY, Notkestra{\ss}e  85,  22607  Hamburg, Germany}

\author{Herbert K. Dreiner} \email{dreiner@uni-bonn.de}
\affiliation{\AddrBonn}


\author{V\'ictor Mart\'in Lozano} \email{victor.lozano@desy.de} 
\affiliation{\AddrBonn,\\ and \AddrDESY}

\begin{abstract}
We consider pair production of LSP staus at the LHC within R-parity violating supersymmetry. The staus decay into Standard 
Model leptons through the $LL\bar{E}$ operator. Using \texttt{CheckMATE} we have recast multileptonic searches to test such 
scenarios. We show for the first time that using these analyses the stau mass can be constrained up to 345\,GeV, 
depending on the stau decay mode, as well as the stau mixing angle. However, there is for all scenarios a 
significant gap between the lower LEP limit on the stau mass and the onset of the LHC sensitivity. This approach can be 
used in the future to constrain the stau sector in the context of RPV lepton-number violating models.
\end{abstract}

\maketitle

\section{Introduction}

Supersymmetry \cite{Nilles:1983ge,Martin:1997ns} is a widely considered possible solution to the hierarchy problem
\cite{Gildener:1976ai,Veltman:1980mj}. When extending the Poincar\'e and gauge symmetries of the Standard Model
of particle physics (SM) to include supersymmetry, in the minimal version, the particle content must be doubled, 
matching fermions with bosons. Furthermore an extra Higgs doublet is added. The most general renormalizable 
superpotential with this field content is
\begin{eqnarray}
W_{\mathrm{SSM}}&=& W_{\mathrm{MSSM}}+W_{\mathrm{RPV}}\,, \\[2mm]
W_{\mathrm{MSSM}}&=& \epsilon_{ab}\left[h^E_{ij}L_i^aH_d^b\bar{E}_j + h^D_{ij}Q_i^aH_d^b\bar{D}_j +h^U_{ij}
Q_i^aH_u^b\bar{U}_j 
\right.  \notag\\[2mm]
&&\left.  +\mu H_d^aH^b_u\right]\,,\label{eq:W-MSSM}\\
W_{\mathrm{RPV}}&=&\epsilon_{ab}\left[\frac{1}{2}\lambda_{ijk}L_i^aL_j^b\bar{E}_k + \lambda'_{ijk}L_i^aQ_j^b
\bar{D}_k-\kappa_iL_i^aH^b_u
\right] \notag \\&&+\dfrac{1}{2}\epsilon_{xyz}\lambda''_{ijk} \bar{U}_i^x\bar{D}_j^y\bar{D}_k^z \,.
\label{eq:W-RPV}
\end{eqnarray}
Here we have used the common notation employing chiral superfields of for example 
Ref.~\cite{Allanach:1999mh,Allanach:2003eb}. The operators in Eq.~(\ref{eq:W-MSSM}) lead to masses for the SM 
fermions and mixing in the Higgs sector. The operators in the first line of Eq.~(\ref{eq:W-RPV}) violate lepton-number, 
those in the second line violate baryon-number.  Together these latter operators lead to a proton decay rate in 
disagreement with the experimental limits, unless the couplings are extremely small, see for example
Ref.~\cite{Smirnov:1996bg}. In the case of the MSSM (minimal supersymmetric Standard Model) the discrete 
multiplicative symmetry R-parity is imposed, where
\begin{equation}
R_p=(-\mathbf{1})^{2S+3B+L}\,,
\end{equation}
with $S$ the spin, $B$ the baryon-number and $L$ the lepton-number of a particle \cite{Farrar:1978xj}. This prohibits 
all the baryon- and lepton-number violating operators in Eq.~(\ref{eq:W-RPV})  and the proton is stable. Furthermore, 
in such R-parity conserving supersymmetric models (RPC) the lightest supersymmetric particle (LSP) is stable. For 
cosmological reasons it must be electrically neutral \cite{Ellis:1990nb} and is usually considered to be the lightest 
neutralino. It has been extensively studied as a dark matter candidate \cite{Goldberg:1983nd}. The RPC model must be 
extended for example by a heavy see-saw sector to allow for light neutrino masses. R-parity is discrete gauge 
anomaly-free \cite{Ibanez:1991hv}, however it allows for dimension-five proton decay operators. Thus the symmetry 
proton hexality $P_6$, is preferable, which at colliders is phenomenologically equivalent \cite{Dreiner:2005rd}.

Models where a subset of the terms in Eq.~(\ref{eq:W-RPV}) are allowed are called R-parity violating supersymmetric 
models, short RPV models \cite{Dreiner:1997uz,Allanach:2003eb,Barbier:2004ez}. For example with the discrete 
symmetry baryon triality \cite{Dreiner:2006xw,Dreiner:2011ft} only the lepton-number violating couplings are allowed. 
Just as R-parity, baryon triality is discrete gauge anomaly-free \cite{Ibanez:1991hv,Dreiner:2005rd,Dreiner:2012ae}, 
and can thus be 
consistently embedded in higher energy models without violation through quantum gravity effects. From a theoretical 
point of view RPV models are thus at least as well motivated as R-parity conserving models. As a benefit, light neutrino 
masses are obtained automatically \cite{Hall:1983id,Hempfling:1995wj,Dreiner:2006xw,Dreiner:2011ft}, notably 
\textit{without} an additional heavy Majorana neutrino scale.

Due to the terms in $W_{\mathrm{RPV}}$ the neutralino LSP is no longer stable and is not a dark matter candidate. 
Instead,  a potential dark matter candidate is the axino 
\cite{Chun:1999cq,Choi:2001cm,Colucci:2018yaq,Colucci:2015rsa,Colucci:2018yaq}, which is also unstable, but due 
to the small coupling is long lived on cosmological time scales. We shall not further consider the axino here as it 
is irrelevant for collider physics. We denote as the LSP, the lightest non-axino supersymmetric particle.

Since the LSP is not constrained by cosmological considerations, in principle any supersymmetric particle (sparticle) 
can be the LSP.  Allowing for any sparticle to be the LSP leads to a very wide range of potential signatures at the LHC, 
many dramatically different from the standard missing transverse momentum signatures in RPC 
\cite{Dreiner:1991pe,Allanach:2006st,Dercks:2017lfq}. If we assume a simple set of boundary conditions for the 
supersymmetric parameters at the unification scale $M_X=\mathcal{O}(10^{16}\,\mathrm{GeV})$, \textit{i.e.} the 
constrained minimal supersymmetric Standard Model (CMSSM) \cite{Martin:1997ns,Bechtle:2013mda,Bechtle:2015nua} 
modified by an additional RPV operator, and run the spectrum down to the weak scale via the renormalization group
equations (RGEs), including the RPV couplings \cite{Allanach:1999mh,Allanach:2003eb,Allanach:2006st}, then only a 
small set of sparticles can be realized as the LSP 
\cite{Dreiner:2008ca,Dercks:2017lfq}. Already in the R-parity conserving CMSSM the stau is the LSP in large regions of 
parameter space, namely when $M_0\lsim M_{1/2}/6$ \cite{Allanach:2006st,Dreiner:2008ca}. As mentioned, these are 
excluded for cosmological reasons in RPC \cite{Ellis:1990nb}, however in RPV models they are allowed. Thus even for 
very small RPV-couplings the stau can be the LSP. These parameter ranges are extended in $M_0,\,M_{1/2}$ for larger 
values of RPV-couplings involving the stau, in particular the right-handed stau, \textit{i.e.} $\lam_{ij3}$, see Fig.~2 in 
Ref.~\cite{Dercks:2017lfq}.

In Ref.~\cite{Dercks:2017lfq} the coverage of the RPV CMSSM with a stau LSP through existing LHC data was 
investigated, and found to be almost non-existent \cite{Desch:2010gi,ATLAS:2012kr}. A specific stau-LSP benchmark 
point is given in Ref.~\cite{Allanach:2006st}. See also the section on supersymmetric particle searches in the PDG 
\cite{Tanabashi:2018oca}. It is the purpose of this paper to investigate the phenomenology of supersymmetric RPV 
stau-LSP models at the LHC. To be definite, we focus on solely  non-zero $LL\bar E$ operators, for which the stau 
decays directly via a two-body mode, \textit{i.e.} $L_iL_j\bar E_3,$ or $L_iL_3\bar E_{1,2}$, with $i,j\in\{1,2,3\}$. The 
outline of this paper is as follows. In Sec.~\ref{sec:model} we present our model, as well as the specific 
representative scenarios with their LHC signatures, which we investigate in detail. In Sec.~\ref{sec:stau-decays} we 
discuss the stau decay branching ratios as well as the stau lifetime. In Sec.~\ref{sec:LEP-Bounds} we review the
RPV stau searches at LEP, in particular the resulting lower mass bounds. In Sec.~\ref{sec:recasting} we discuss the 
experimental LHC analyses we employ in recasting. In Sec.~\ref{sec:numerics} we present our numerical results and in 
Sec.~\ref{sec:conclusions} we offer our conclusions.

\section{Model}
\label{sec:model}
In the minimal supersymmetric Standard Model, the off-dagonal elements in the left-right (LR) single flavor  sfermion 
mass matrices are proportional to the fermion mass. Thus for sleptons, the stau will have the largest mixing. After 
diagonalizing the mass matrix (see for example Ref.~\cite{Drees:1996ca}), we obtain the mass eigenstates $\tilde
\tau_{1,2}$ in terms of the SU(2)$_L$ current eigenstates $\tilde\tau_{L,R}$
\begin{equation}
\left(
\begin{array}{c}
\tilde{\tau}_1 \\
\tilde{\tau}_2
\end{array}
\right) =
\left(
\begin{array}{cc}
\cos \theta_{\tilde{\tau}} & \sin \theta_{\tilde{\tau}} \\
-\sin \theta_{\tilde{\tau}}&\cos \theta_{\tilde{\tau}}
\end{array}
\right)
\left(
\begin{array}{c}
\tilde{\tau}_L \\
\tilde{\tau}_R
\end{array}
\right).
\label{eq:stau-mixing}
\end{equation}
Here for the masses: $m_{\tilde\tau_1}< m_{\tilde\tau_2}$, and $\theta_{\tilde\tau}$ is the mixing angle in the 
stau sector. Together with the stau mass it is the main free parameter in our analysis. For $\theta_{\tilde\tau}=0$ the 
lightest stau, $\tilde\tau_1$, is pure $\tilde\tau_L$, for $\theta_{\tilde\tau}=\frac{\pi}{2}$ it is pure $\tilde\tau_R$. In the 
following we consider the lightest stau to be the LSP and analyze direct pair production at the LHC
\begin{equation}
pp\to\tilde\tau^+_1\tilde\tau^-_1+X\,.
\end{equation}
Here we only consider the production of the staus via gauge couplings, \textit{i.e.} we consider the RPV couplings to be 
small compared to the gauge couplings, in accordance with the present limits 
\cite{Barbier:2004ez,Kao:2009fg,Dreiner:2012mx}. (For larger couplings single sparticle production is more promising
\cite{Dreiner:2000vf,Dreiner:2006sv,Dreiner:2012np}.) R-parity violation leads to the stau decaying in the detector, if the 
RPV coupling is not too small. We discuss this in detail below. Depending on the decay mode, we focus on three 
different models, each with only one dominant RPV operator, and for which the stau decays as 
\begin{eqnarray}
\mathbf{Model\,I:}\;& \;L_aL_3\bar E_c,\;\;& \,\tilde\tau_1^+\to \ell_c^+\nu_a, \label{eq:Model-1}
\\[1mm]
\mathbf{Model\,II:}\;&\;L_1L_2\bar E_3,\;\;& \,\tilde\tau_1^+\to (e^+\bar\nu_\mu,\mu^+\bar\nu_e), \label{eq:Model-2}\\[1mm]
\mathbf{Model\,III:}\;&\;L_aL_3\bar E_3,\;\;& \,\tilde\tau_1^+\to (\tau^+\nu_a,\tau^+\bar\nu_a,\ell^+_a\nu_\tau),\;\; \label{eq:Model-3}
\end{eqnarray}
where $a,c\in\{1,2\}$, and the $\tilde\tau^-_1$ decay to the charge conjugate final states. 

Out of Models I-III in Eqs.~(\ref{eq:Model-1})-(\ref{eq:Model-3}), we consider 5 separate scenarios, which are listed in 
Tab.~\ref{tab:signatures}. We consider Model I with $c=1$ as Model Ia, and with $c=2$ as Model Ib. Similarly we shall 
consider Model III separately with $a=1$, Model IIIa, and with $a=2$, Model IIIb. This corresponds to treating electrons 
and muons separately.
\begin{table}[t]
\begin{tabular}{cccc}
Model\; & Coupling & \;$\tilde\tau_1$-Decays\; & \;\;Signatures \\ \hline
Ia & $\lam_{a31}$ & $e\nu_a$& \;$e^+e^-+\not \!\!E_T$  \\[1.5mm]
Ib & $\lam_{a32}$ & $\mu\nu_a$& \;$\mu^+\mu^-+\not \!\!E_T$  \\ [1.5mm]
II & $\lam_{123}$ & $\mu\nu_e,e\nu_\mu$& \;$e^+e^-+\not \!\!E_T$  \\ [1.mm]
&&& \;$\mu^+\mu^-+\not \!\!E_T$  \\ [1mm]
&&& \;$e^\pm\mu^\mp+\not \!\!E_T$  \\ [1.5mm]
IIIa & $\lam_{133}$ & $e\nu_\tau$,$\tau\nu_e$& \;$e^+e^-+\not \!\!E_T$  \\ [1mm]
&& & \;$\tau^+\tau^-+\not \!\!E_T$  \\ [1mm]
&& & \;$e^\pm\tau^\mp+\not \!\!E_T$  \\ [1.5mm]
IIIb & $\lam_{233}$ & $\mu\nu_\tau$,$\tau\nu_\mu$& \;$\mu^+\mu^-+\not \!\!E_T$  \\ [1mm]
&& & \;$\tau^+\tau^-+\not \!\!E_T$  \\ [1mm]
&& & \;$\mu^\pm\tau^\mp+\not \!\!E_T$  
\end{tabular}
\caption{Signatures for the pair production of staus at the LHC, with two-body decays of the staus via the 
$LL\bar E$ operators.  Column 2 lists the dominant single coupling we consider, column and in particular those leading to two-body decays of the staus, column 2. This leads to five models,
when considering $e^\pm$ and $\mu^\pm$ separately, column 1. In column 3 we list the decay modes of the $\tilde\tau_1$ 
and in column 4 the signatures for a pair of staus. Here the generation index $a=1,2$.}
\label{tab:signatures}
\end{table}

Below we discuss each scenario in detail. We recast them in terms of LHC searches implemented in the program
\texttt{CheckMATE}  \cite{Drees:2013wra,Dercks:2016npn}. As we see below, the most relevant searches are the ones with leptons 
and missing energy in the final state, which are dedicated to the search for electroweakinos or stop squarks.


\section{Stau Decays}
\label{sec:stau-decays}
\subsection{Branching Ratios}

According to Eq.~(\ref{eq:stau-mixing}), the lightest stau eigenstate is given by the mixture
\begin{equation}
\tilde\tau_1= \cos\theta_{\tilde\tau} \,\tilde\tau_L +\sin\theta_{\tilde\tau} \cdot\tilde\tau_R \,.
\label{eq:mixing-stau}
\end{equation}
In Model I only the $\tilde\tau_L$-component of $\tilde\tau_1$ couples to the dominant operator. Furthermore there is only one
two-body decay mode. In this case the $\tilde\tau_1$ decay rate is given by \cite{Richardson:2000nt}
\begin{equation}
\Gamma(\tilde\tau_1^+\to\ell^+_c\nu_a)=|\lam_{a3c}|^2 \cos^2{\theta_{\tilde\tau}}\,
\frac{(m_{\tilde\tau_1}^2-m_{\ell_c}^2)^2}{16\pi m_{\tilde\tau_1}^3}\,,\label{eq:gammamodeli}
\end{equation}
where $\ell^+_c$ denotes the final state charged lepton of generation $c=1,2$, and and $m_{\ell_c}$ its mass. For $\theta_{\tilde
\tau}\to\frac{\pi}{2}$ this decay width vanishes. The stau then decays to a four-body final state 
\cite{Allanach:2003eb,Allanach:2006st} via a virtual neutralino
\begin{equation}
\tilde\tau_1^-\to\tau^- +(\tilde\chi^0_1)^* \to \tau^- +(\tau^\pm \ell_c^\mp\nu_a,\,\ell_a^\pm\ell_c^\mp\nu_\tau)\,,
\end{equation}
with a total of four decay modes. The decay rate is given in the appendix of 
Ref.~\cite{Allanach:2003eb}.  For fixed slepton and neutralino mass the partial width goes as $m_{\tilde\tau_1}^7$. A 
related decay via the chargino is also possible,
\begin{equation}
\tilde\tau_1^-\to\nu_\tau +(\tilde\chi^-_1)^* \to \nu_\tau+(\ell_a^\pm \tau^\mp\ell_c^\mp,\nu_a\nu_\tau\ell_c^-)\,,
\end{equation}
with a similar decay rate. We have assumed here that the lightest chargino is wino dominated.

\begin{figure}[t]
	\begin{center}
		\includegraphics[width=.85\linewidth]{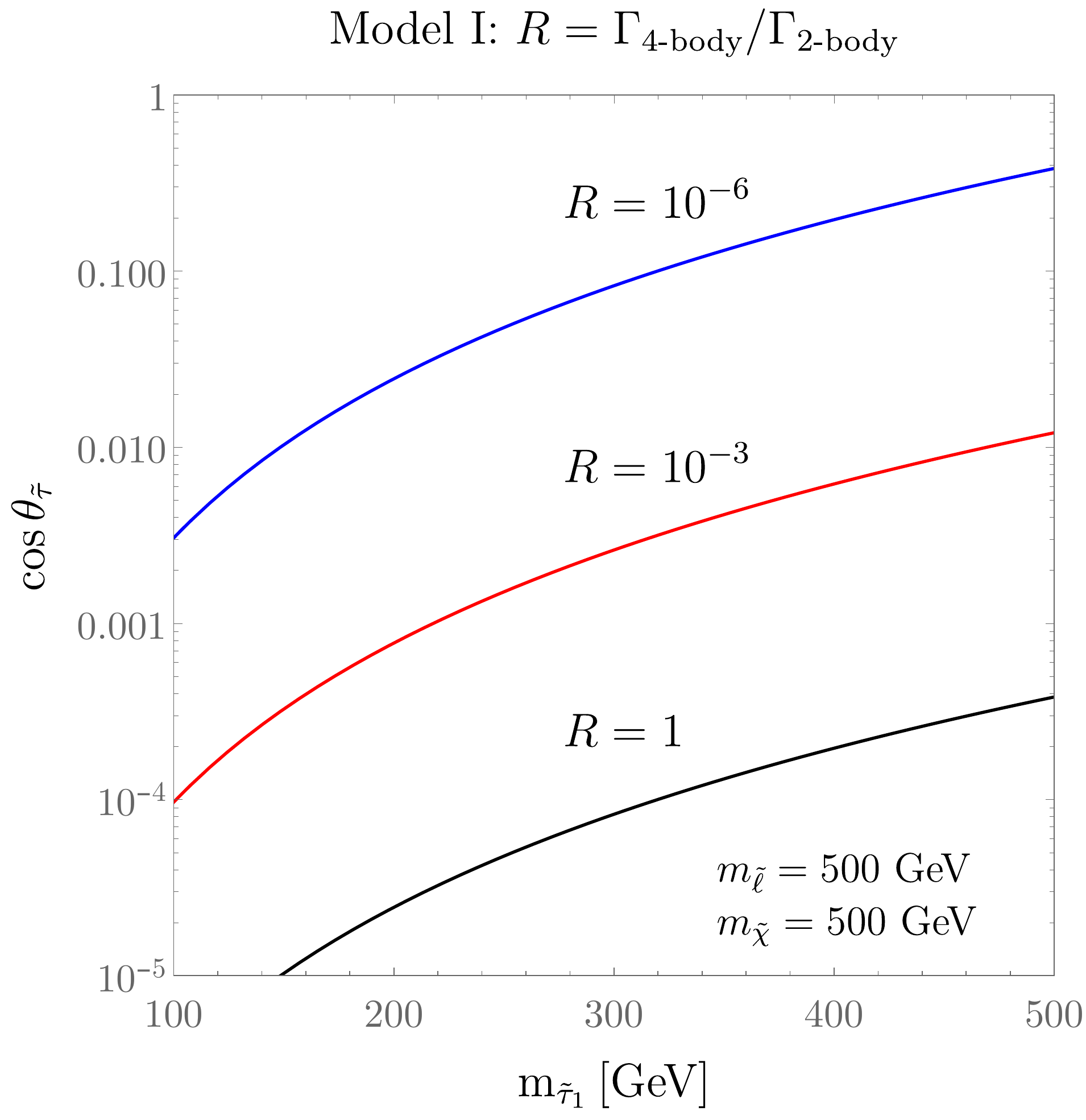}
	\end{center}
	\caption{Isocurves of the ratio of the four-body and the two-body stau decay widths as a function of the stau mass and 
	$\cos\theta_{\tilde\tau}$ in Model I for masses of the intermediate particles $m_{\tilde{\chi}}=m_{\tilde{\ell}}=500\,$GeV.
	For the neutralino for simplicity, we have assumed SU(2)$_L$ couplings only.}
	\label{fig:stau-decays-4vs2-modI}
\end{figure}

In Fig.~\ref{fig:stau-decays-4vs2-modI} we show isocurves of $R\equiv{\Gamma_{\mathrm{4-body}}}/{\Gamma_{\mathrm{2-body}}}$,
the ratio of the four-body stau decay width over the two-body decay width as a function of the stau mass, $m_{{\tilde \tau}_1}$, and 
$\cos\theta_{\tilde\tau}$ for intermediate masses of $m_{\tilde{\chi}}=m_{\tilde{\ell}}=500\,$GeV. These intermediate masses are 
chosen to maximize the four-body partial decay width. For the neutralino for simplicity, we have assumed SU(2)$_L$ couplings only.
The ratio $R$ grows with $m_{\tilde\tau_1}$ as expected, but remains smaller than about 10$^{-3}$ except for the mixing angle
very close to $\pi/2$. Thus for most of the parameter region the two-body decays are sufficient to understand the results. However, we 
include the four-body decays in our complete analysis.

\begin{figure}[t]
	\begin{center}
		\includegraphics[width=.85\linewidth]{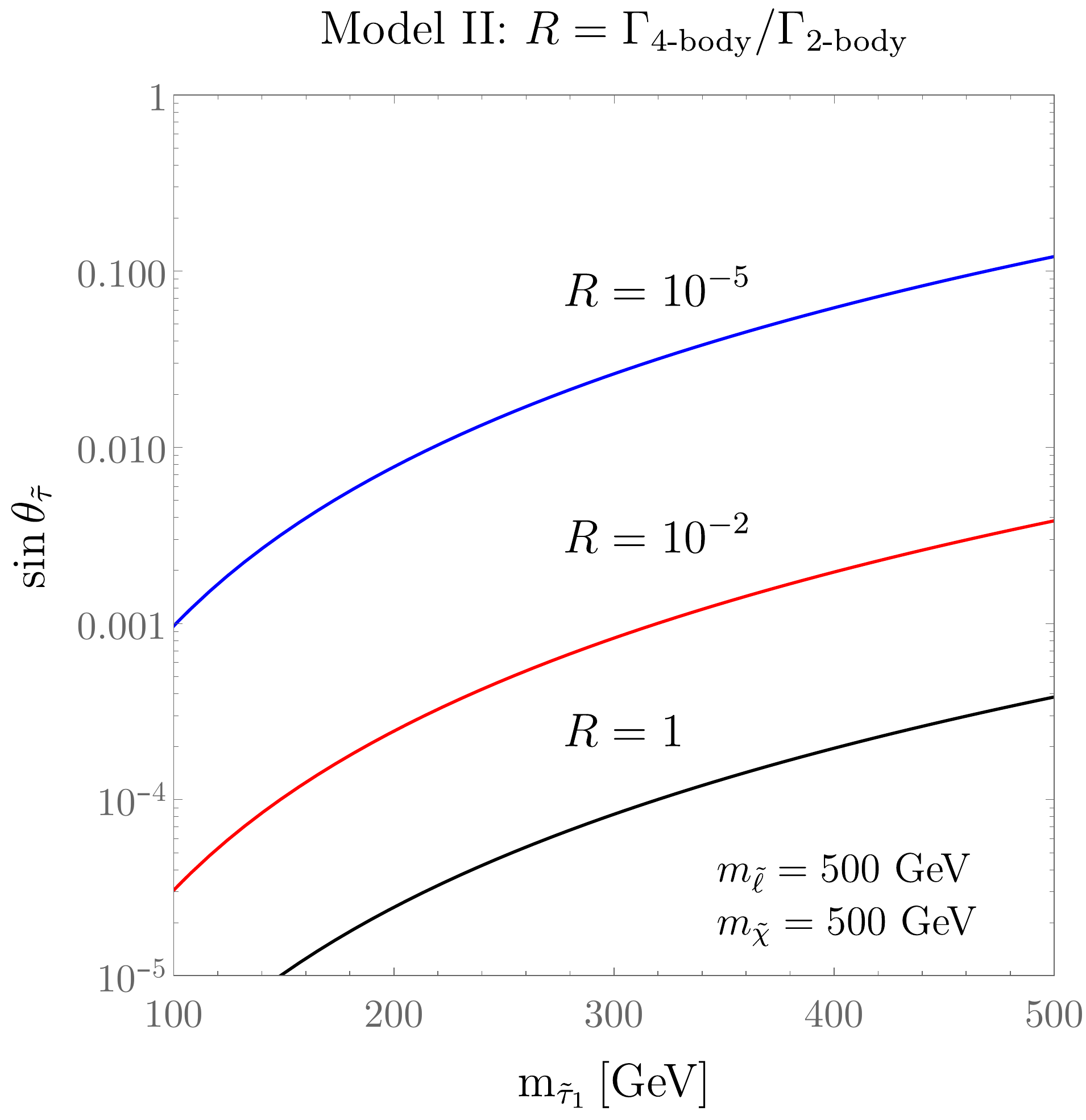}
	\end{center}
	\caption{Isocurves of the ratio of the two-body and the two-body stau decay widths as a function of the stau mass and 
		$\cos\theta_{\tilde\tau}$ in Model II for masses of the intermediate particles $m_{\tilde{\chi}}=m_{\tilde{\ell}}=500\,$GeV.}
	\label{fig:stau-decays-4vs2-modII}
\end{figure}

In Model II, it is only the $\tilde\tau_R$-component of $\tilde\tau_1$ which couples directly to the RPV operator, and there 
are two decay modes. One partial decay width is given by
\begin{equation}
\Gamma(\tilde\tau_1^+\to e^+\bar\nu_\mu)=|\lam_{123}|^2 \sin^2{\theta_{\tilde\tau}}\,
\frac{(m_{\tilde\tau_1}^2-m_e^2)^2}{16\pi m_{\tilde\tau_1}^3}\,. \label{eq:gammamodelii}
\end{equation}
For the other decay rate, $\Gamma(\tilde\tau_1\to\mu^+\bar\nu_e)$, replace $m_e\to m_\mu$.
Neglecting the electron and muon masses, the rates for these two decays in Model II are equal. The branching ratios are 50\%,
respectively, if there are no further decay modes. For $\theta_{\tilde\tau}\to 0$ the decay rate vanishes and the corresponding 
four-body decay modes via virtual electroweakinos must be included. In Fig.~\ref{fig:stau-decays-4vs2-modII} we 
plot isocurves of the ratio $R$ as a function $m_{\tilde\tau_1}$ and $\sin\theta_{\tilde\tau}$. In this model $R>0.01$ for $\theta_{
\tilde\tau}>5\cdot10^{-3}$, and the four-body decays can mostly be neglected.

\begin{figure}[t]
	\begin{center}
		\includegraphics[width=.85\linewidth]{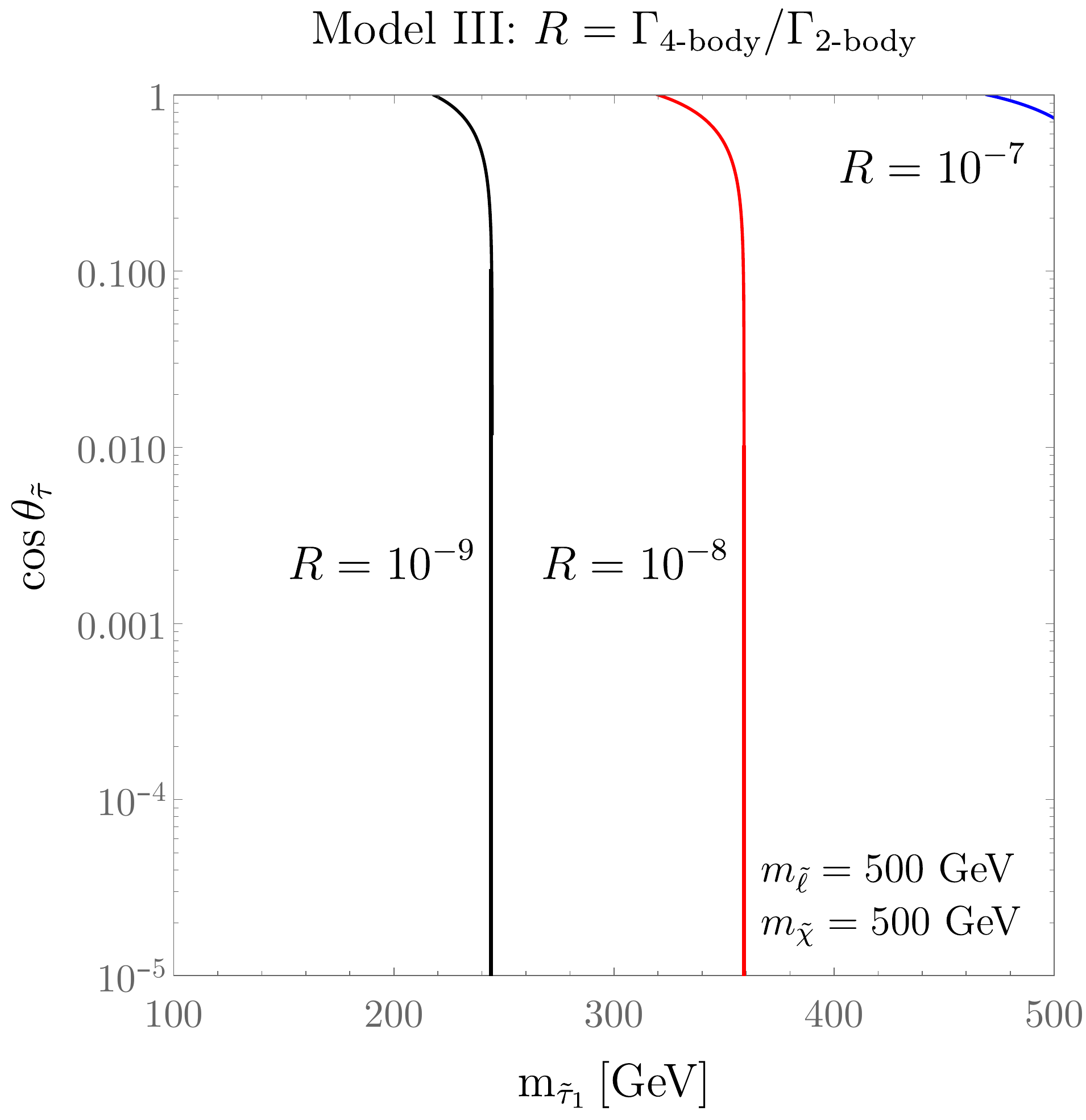}
	\end{center}
	\caption{Isocurves of the ratio of the four-body and the two-body stau decay widths as a function of the stau mass and 
		$\cos\theta_{\tilde\tau}$ in Model III for masses of the intermediate particles $m_{\tilde{\chi}}=m_{\tilde{\ell}}=500\,$GeV.}
	\label{fig:stau-decays-4vs2-modIII}
\end{figure}

In Model III, both the $\tilde\tau_R$ and the $\tilde\tau_L$ components separately couple to the dominant operator. The partial widths 
for the three decay modes are
\begin{eqnarray}
\Gamma(\tilde\tau_1^+\to \tau^+\bar\nu_a)&=&|\lam_{a33}|^2 \sin^2{\theta_{\tilde\tau}}\,
\frac{(m_{\tilde\tau_1}^2-m_\tau^2)^2}{16\pi m_{\tilde\tau_1}^3}\,, \label{eq:stau1}\\
\Gamma(\tilde\tau_1^+\to \ell^+_a\bar\nu_3)&=&|\lam_{a33}|^2 \sin^2{\theta_{\tilde\tau}}\,
\frac{(m_{\tilde\tau_1}^2-m_{\ell_a}^2)^2}{16\pi m_{\tilde\tau_1}^3}\,, \label{eq:stau2}\\
\Gamma(\tilde\tau_1^+\to \tau^+\nu_a)&=&|\lam_{a33}|^2 \cos^2{\theta_{\tilde\tau}}\,
\frac{(m_{\tilde\tau_1}^2-m_\tau^2)^2}{16\pi m_{\tilde\tau_1}^3}\,. \label{eq:stau3}
\end{eqnarray}
Neglecting the final-state charged lepton masses compared to the stau mass we thus have for the total width
\begin{equation}
\Gamma(\tilde\tau_1^+\to \tau^+\nu_a)=\frac{|\lam_{a33}|^2}{16\pi}
 (1+\sin^2{\theta_{\tilde\tau}})\,
m_{\tilde\tau_1}\,,\label{eq:gammamodeliii}
\end{equation}
which is non-zero for all values of the stau mixing angle. Therefore $R\ll1$ over the entire parameter range we consider here,
as can be seen in Fig.~\ref{fig:stau-decays-4vs2-modIII}.

Assuming that only the two-body decays are dominant and combining the two decays to $\tau^+$'s, Eqs.~(\ref{eq:stau1}) and 
(\ref{eq:stau3}), which are observationally equivalent, we obtain for the pure two-body branching ratios
\begin{eqnarray}
\mathrm{Br}(\tilde\tau_1\to\{\tau^+\bar\nu_a,\tau^+\nu_e\})&=& \frac{1}{1+\sin^2\theta_{\tilde\tau}}\,, \label{eq:stau-Br1}\\
\mathrm{Br}(\tilde\tau_1\to \ell_a^+\bar\nu_\tau) &=& \frac{\sin^2\theta_{\tilde\tau}}{1+\sin^2\theta_{\tilde\tau}}\,.
\label{eq:stau-Br2}
\end{eqnarray}
Thus for $\cos\theta_{\tilde\tau}\to1$ the stau decays 100\% to $\tau$-leptons, which is important for searches. The maximum
branching ratio to $\ell_a^+\in\{e^+,\mu^+\}$ is only 50\%, obtained for $\cos\theta_{\tilde\tau}\to0$.

\subsection{Stau Decay length}
We next consider an estimate for the stau LSP lifetime. Ignoring the final state charged lepton masses, using only the two-body 
decay width formula, and setting $m_{\tilde\tau_1}=250\,$GeV we estimate
\begin{equation}
\tau_{\tilde\tau_1}=\frac{1.3\cdot10^{-25}\,\mathrm{s}} {\lam_{ijk}^2\cdot A_{\theta_{\tilde\tau}}^2}
\end{equation}
where 
$A_{\theta_{\tilde\tau}}^2\equiv[\mathrm{I\!:}\, \cos^2{\theta_{\tilde\tau}};\,\mathrm{II\!:}\, \sin^2{\theta_{\tilde\tau}};\,
\mathrm{III\!:}\,(1+\sin^2{\theta_{\tilde\tau}})]$. Thus for 
\begin{equation}
\lam_{ijk}\cdot A_{\theta_{\tilde\tau}} \gsim 2\cdot 10^{-7}\label{eq:limitctau}
\end{equation}
the stau decays promptly in the detector, $c\tau_{\tilde\tau_1}\lsim1\,$mm. Depending on the stau mixing angle, for a wide range
of couplings, $\lam_{ijk}$, consistent with existing upper bounds 
\cite{Dreiner:1992vm,Allanach:1999ic,Barbier:2004ez,Kao:2009fg,Dreiner:2010ye}, we can 
consider the staus to decay promptly.\footnote{In related recent work \cite{Bansal:2019zak} on RPV tau physics at the LHC, bounds 
were set on the $\lam'$ couplings in $LQ\bar D$ scenarios, which we shall discuss elsewhere.} 


\begin{figure}[t]
	\begin{center}
		\includegraphics[width=.85\linewidth]{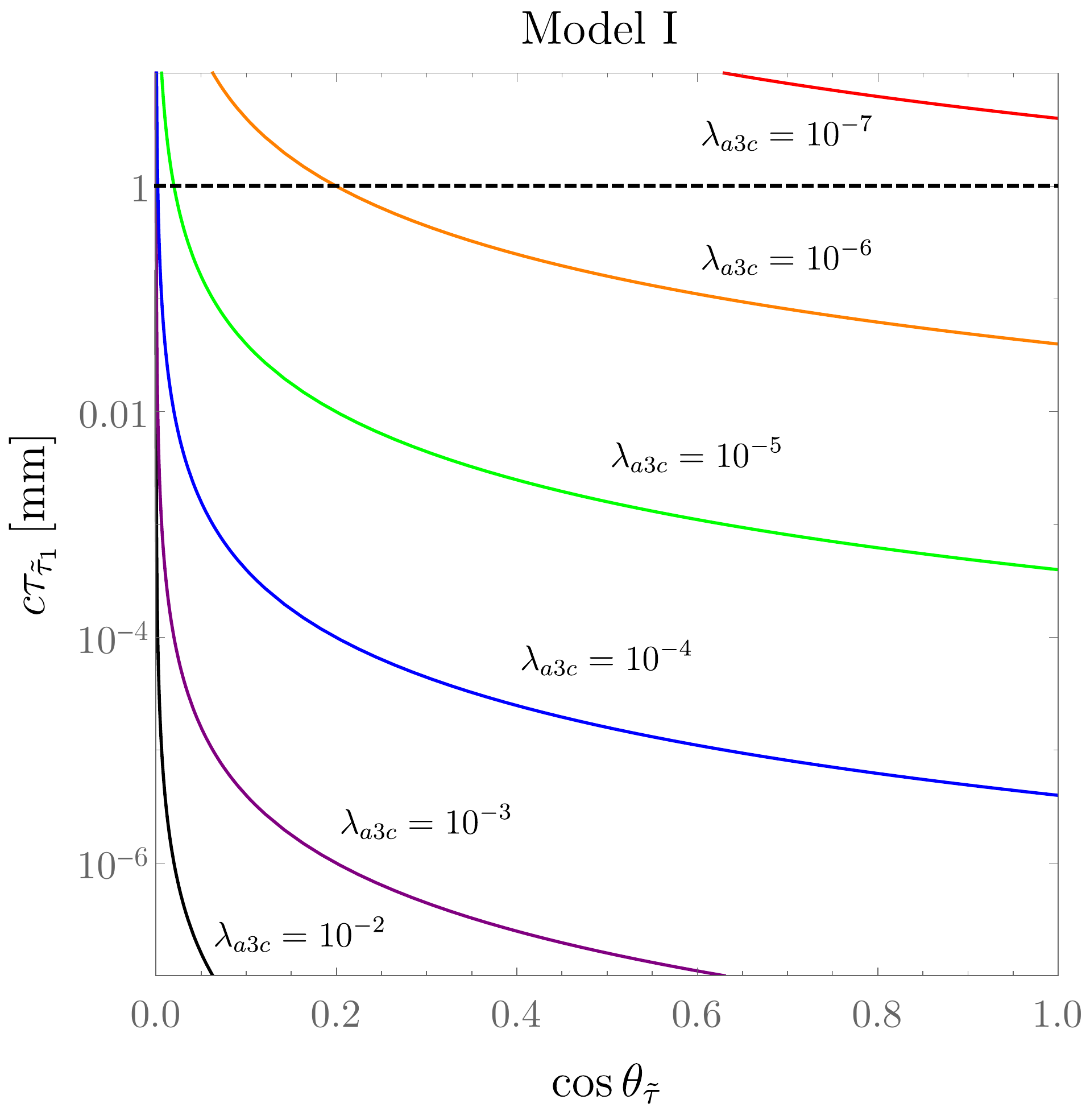}
	\end{center}
	\caption{Decay length of the stau LSP, $\tilde{\tau}_1$, as a function of the mixing angle, $\cos \theta_{\tilde \tau}$, for different 
	values of the $\lambda_{a3c}$ coupling.}
	\label{fig:stau-decays-dlength-modI}
\end{figure}

We performed the complete computation to check Eq.~\eqref{eq:limitctau} quantitatively. In Fig.~\ref{fig:stau-decays-dlength-modI} 
we plot the decay length for $m_{\tilde{\tau}_1}=250\,$GeV, as a function of the mixing angle, $\cos\theta_{\tilde \tau}$, for different
values of the coupling $\lam_{a3c}$, \textit{i.e.} Model I. The horizontal dashed line denotes $c\tau_{\tilde\tau_1}=1\,$mm, below 
which we consider the decay prompt. When the mixing angle approaches $\pi/2$, \textit{i.e.} $\cos\theta_{\tilde \tau}\to 0$, the 
decay length grows, as expected from Eq.~\eqref{eq:gammamodeli}, but then the four-body decays kick in. In this extreme
case the stau becomes long-lived for $\lam_{a3c}\lsim 3\cdot10^{-7}$. For  $\lam_{a3c}\lesssim 10^{-5}$ the stau becomes long-lived 
for non-negligible values of $\cos\theta _{\tilde \tau}$. In particular, for $\lam_{a3c}=10^{-6}$, we have  $c\tau_{\tilde\tau_1} > 1\,$mm for $\cos\theta_{\tilde \tau}< 0.174$.  For $\lam_{a3c}<3\cdot10^{-7}$, we find $c\tau\geq 1\,$mm independently of the mixing angle.

\begin{figure}[t]
	\begin{center}
		\includegraphics[width=.85\linewidth]{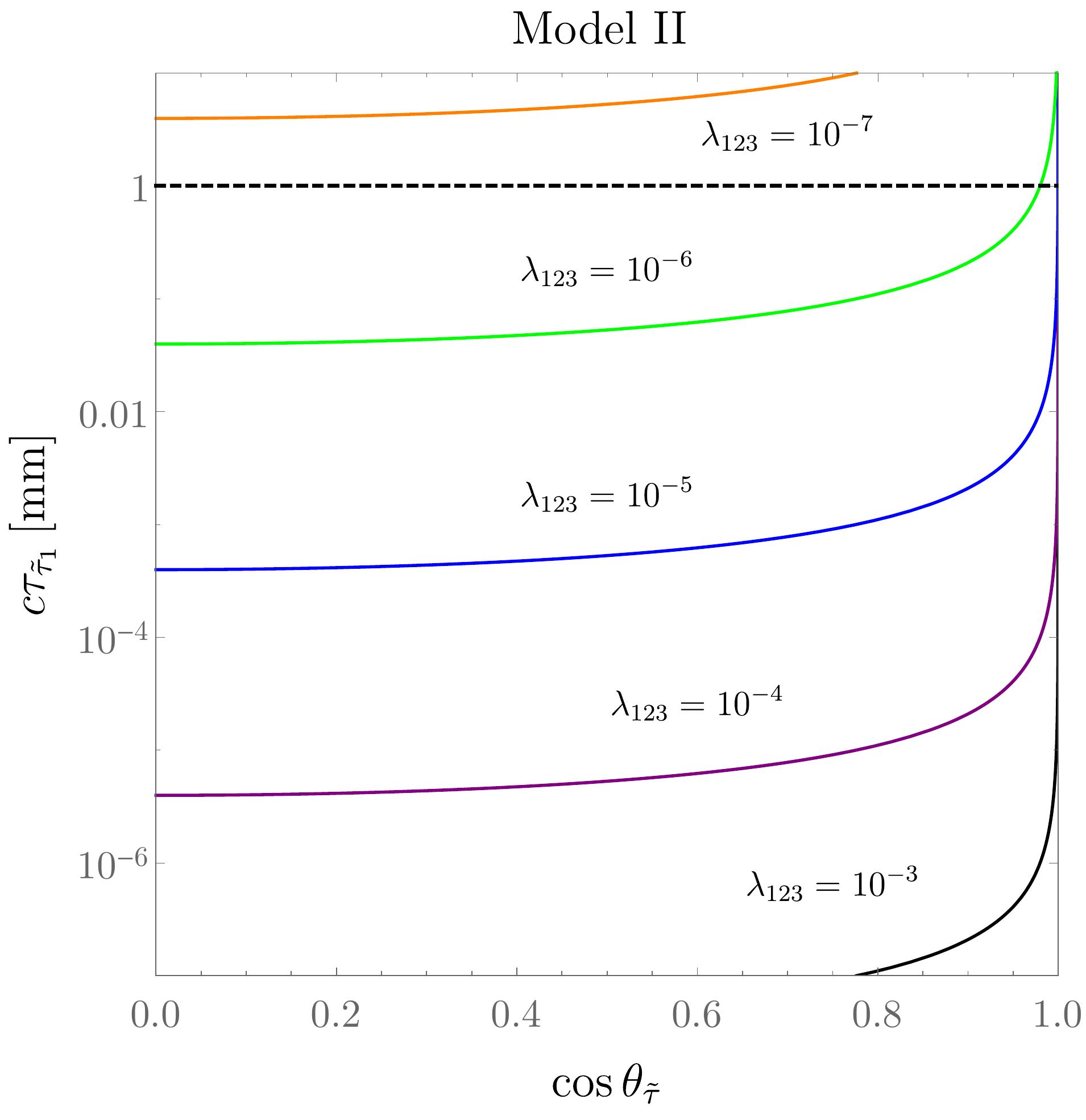}
	\end{center}
	\caption{Decay length of the stau LSP, $\tilde{\tau}_1$, as a function of the mixing angle, $\cos \theta_{\tilde \tau}$, for different values of the $\lambda_{123}$ coupling.}
	\label{fig:stau-decays-dlength-modII}
\end{figure}

\begin{figure}[t]
	\begin{center}
		\includegraphics[width=.85\linewidth]{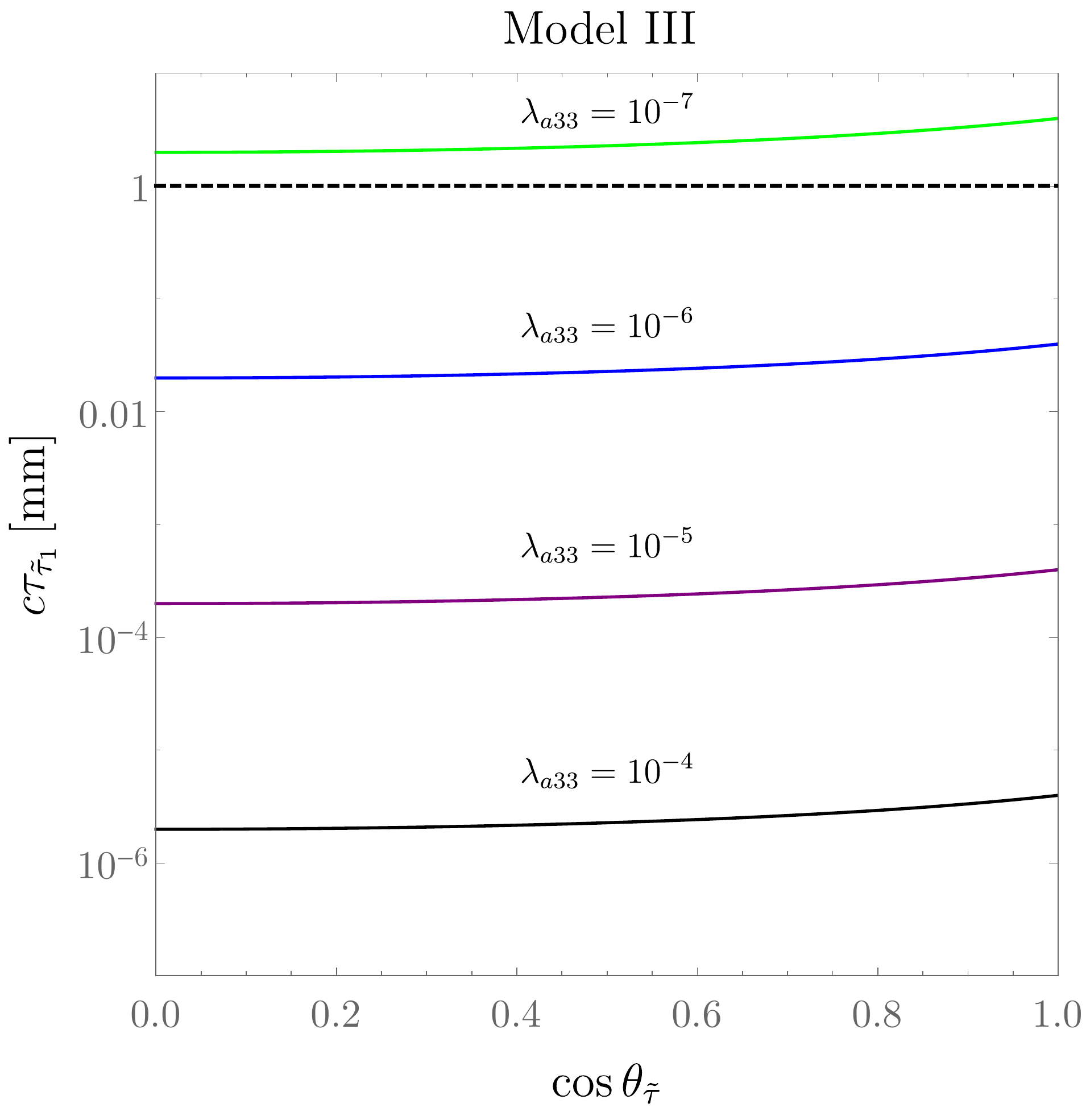}
	\end{center}
	\caption{Decay length of the stau LSP, $\tilde{\tau}_1$, as a function of the mixing angle, $\cos \theta_{\tilde \tau}$, for different values of the $\lambda_{a33}$ coupling.}
	\label{fig:stau-decays-dlength-modIII}
\end{figure}

In Model II we show in Fig.~\ref{fig:stau-decays-dlength-modII} the stau decay length  as a function of the mixing angle, $\cos 
\theta_{\tilde\tau}$, for different values of the parameter $\lam_{123}$. Unlike Model I, the stau becomes long-lived for mixing 
angles close to 0 ($\cos\theta_{\tilde \tau}\to 1$), \textit{cf.} Eq.~\eqref{eq:gammamodelii}. For $\lam_{123}\lesssim 
1.5\cdot10^{-7}$ the decay length becomes greater than $c\tau\geq 1$ mm, basically for all mixing angles.

The analogous plot for Model III is shown in Fig.~\ref{fig:stau-decays-dlength-modIII}. Here the decay length is practically 
independent of the mixing angle. There is only a moderate increase for $\cos\theta_{\tilde \tau}\to 1$. This is expected from
Eq.~\eqref{eq:gammamodeliii}, due to the term independent of $\theta_{\tilde\tau}$. The decay is prompt for
$\lam_{a33}\gsim1.5\cdot10^{-7}$, for all $\theta_{\tilde\tau}$.

We therefore assume that in every model the parameter $\lambda_{ijk}$ is such that the stau decays promptly. As we have seen 
in Figs.~\ref{fig:stau-decays-dlength-modI}, \ref{fig:stau-decays-dlength-modII} and \ref{fig:stau-decays-dlength-modIII} one can 
assure that this condition is fulfilled when $\lambda_{ijk}\gtrsim 10^{-6}$. \

For $m_{\tilde\tau_1}\gsim150\,$GeV the low-energy
constraints on the R-parity violating couplings are fulfilled for $\lam_{ijk}<0.1$ \cite{Allanach:1999ic, Dercks:2017lfq}, which is 
beyond the range plotted in Figs.~~\ref{fig:stau-decays-dlength-modI}, \ref{fig:stau-decays-dlength-modII} and 
\ref{fig:stau-decays-dlength-modIII}.

\section{Lower LEP Limits on Stau Mass}
\label{sec:LEP-Bounds}
All four experiments at LEP have published papers on searches for staus in RPV supersymmetric scenarios; \texttt{ALEPH}: 
\cite{Heister:2002jc}, \texttt{DELPHI} \cite{Abdallah:2003xc}, \texttt{L3} \cite{Achard:2001ek}, and \texttt{OPAL} \cite{Abbiendi:2003rn}.
We briefly summarize their results here, as we need them below. The experiments do not perform a systematic analysis of the bounds 
for a stau LSP for arbitrary mixing angles, which is what we would need. Most of the analyses are on pure right-handed staus,
since in unification scale models the largest component of $\tilde\tau_1$ is typically $\tilde\tau_R$, with a few mass searches also
for pure $\tilde\tau_L$. We thus consider the lower mass bounds at the limiting cases of mixing, \textit{i.e.} for $\cos\theta_{\tilde\tau}
=0$ and $\cos\theta_{\tilde\tau}=1$ and interpolate these for our results in Figs.~\ref{fig:a31}-\ref{fig:133}, below.

All experiments consider direct two-body decays of the staus via the $LL\bar E$ operators, as well as indirect decays via an 
intermediate neutralino, \textit{i.e.} the chargino decays are mentioned, but not taken into account. Both cases are treated 
separately with 100\% branching ratio, respectively, \textit{i.e.} either with 100\% two-body decays, or 100\% four-body decays.
In the indirect case, the neutralino is assumed to be on-shell, and thus lighter than the stau. This does not correspond to our
scenarios. We consider these searches all the same as constraints on our models, as we believe the essential feature is the
kinematics of  the four-body decay of the stau. This is underlined by the fact, that the bounds depend only weakly on the neutralino 
mass outside the kinematic boundaries. We then employ the following bounds at the limiting values.

{\bf Models Ia,b}: for $\cos\theta=0$ we use the \texttt{DELPHI} and \texttt{OPAL} limits on indirect decays of a $\tilde\tau_R$: $m_{\tilde\tau_R}>92\,$GeV. For $\cos\theta=1$, \texttt{OPAL} have a limit on direct decays of $\tilde\tau_L$: $m_{\tilde\tau_R}>
74\,$GeV, which we employ.

{\bf Models II}:
for $\cos\theta=0$ we use the \texttt{ALEPH} lower mass limit for direct decays of the stau $m_{\tilde\tau_R}> 87\,$GeV. There is 
no limit on the indirect decay of a pure $\tilde\tau_L$. Since the production cross section for $\tilde\tau_L$ is higher than for $\tilde
\tau_R$  we employ the $\tilde\tau_R$ bound, from the previous case in the hope that this conservative:
$m_{\tilde\tau_R}>92\,$GeV.

{\bf Model IIIa}: for this model in both limiting cases, $\cos\theta_{\tilde\tau}=0$ or 1, we have possible two-body direct decays.
Thus in both cases we use the direct limits: $m_{\tilde\tau_R}>92\,$GeV, $m_{\tilde\tau_L}>74\,$GeV.

The red regions in Figs.~\ref{fig:a31}-\ref{fig:133}, below, are the interpolation of these bounds. We encourage the experiments
to go back and reanalyze the LEP data in lieu of the models we consider here.

\section{Recasting}
\label{sec:recasting}

We now wish to compare the predictions of the models discussed in the previous section to LHC data. For this we recast the
simulated model results in terms of existing LHC analyses. In order to study every scenario we have produced benchmark points 
making use of the spectrum generator \SPheno~\cite{Porod:2003um,Porod:2011nf,Staub:2017jnp}. For each of these points we 
have generated $2\cdot 10^5$ Monte Carlo (MC) events with the event generator {\tt Pythia\,8.219}~\cite{Sjostrand:2014zea}, 
using the default parton distribution function {\tt NNPDF~2.3}~\cite{Ball:2012cx}. Then we confront the MC events against {\tt 
CheckMATE\,2.0.26}~\cite{Drees:2013wra,Dercks:2016npn} which is based on the fast detector simulation {\tt Delphes\,3.4.0} 
\cite{deFavereau:2013fsa} and the jet reconstruction tool {\tt Fastjet\,3.2.1} \cite{Cacciari:2005hq,Cacciari:2011ma}. {\tt CheckMATE} 
is an analysis tool designed to test models against several ATLAS and CMS searches at 8 and 13~TeV. In order to obtain more 
realistic results we apply some correction factors to the leading order cross section calculated by {\tt Pythia\,8}. For that purpose 
we use the NLO stau production cross section given in Ref.~\cite{Fuks:2013lya} for the case of events produced at $\sqrt{s}=8
$~TeV and Ref.~\cite{Fiaschi:2018xdm} for those at $\sqrt{s}=13$~TeV.\footnote{In Ref.~\cite{Fiaschi:2019zgh} the production 
of sleptons at NNLO+NNLL is considered. However, they found only very moderate increases in the total cross sections compared 
with the NLO + NLL results. Thus we take only the results of the latter.}

In our  grid scans for every scenario, we cover the stau mixing angle range: $\cos \theta_{\tilde \tau}\in [0+\epsilon,1-\epsilon]$,  
with $\epsilon=10^{-3}$, avoiding the long decay length regions. We consider the mass range: $m_{\tilde \tau_1}\in [100, 500]
\,$GeV. Although we have made use of all the ATLAS and CMS searches in {\tt CheckMATE} only a few actually constrain our
scenarios. They are listed in Table~\ref{tab:searches}. As we see, there are three analyses that are important, two from Run~1 and 
one from Run~2. We briefly discuss what makes them relevant for our scenarios. 

\vspace{0.75cm}

\noindent {\bf Search for direct stop production decaying into $2\ell + \met$} (arXiv:1403.4853)~\cite{Aad:2014qaa}: This search 
for stops focuses on leptonic final states, electrons and muons, with opposite charge. The leptons come from the decay of $W^
\pm$ bosons produced in the decay chain of the stop squarks. The two opposite sign $W$'s decay independently, resulting in 
the combinations: $e e$, $e\mu$ and $\mu\mu$. This is important for studying RPV couplings with various flavor indices. This 
search was performed for a center-of-mass energy of $\sqrt{s}=$ 8 TeV and an integrated luminosity of $\mathcal{L}=20.3$~fb$^{-1}$.

\vspace{0.3cm}

\noindent {\bf Search for direct slepton and chargino production decaying into $2\ell + \met$} (ATLAS-CONF-2013-049)~\cite{TheATLAScollaboration:2013hha}: 
The aim of this search is the detection of chargino or slepton pairs through their subsequent decays into leptons and missing energy. The analysis focuses on a 
final state with two leptons and $\met$. This search was performed for a center-of-mass energy of $\sqrt{s}=$ 8 TeV and an integrated luminosity of $\mathcal{L}
=20.3$~fb$^{-1}$.

\vspace{0.3cm}

\noindent {\bf Search for electroweak production of SUSY particles with $2-3\ell + \met$} 
(ATLAS-CONF-2017-039)~\cite{ATLAS:2017uun}: This search focuses on the direct production of charginos and neutralinos and 
their decays into leptons and missing energy. The signature are 2 or 3 leptons in the final state plus missing energy. It 
was performed for a center-of-mass energy of $\sqrt{s}=$ 13 TeV and an integrated luminosity of $\mathcal{L}=36.1$ fb$^{-1}$.

\vspace{0.3cm}

\begin{table}[t]
\begin{tabular}{r|c|c|c}
$\sqrt{s}$& Reference & Final State & $\mathcal{L}[{\rm fb}^{-1}]$\\
\hline
\hline &&&\\[-0.6mm]
8 TeV & 1403.4853~\cite{Aad:2014qaa}  & $2\ell + \met$ & 20.3 \\[1.1mm]
8 TeV & ATLAS-CONF-2013-049~\cite{TheATLAScollaboration:2013hha} & $2\ell + \met$ & 20.3\\[1.1mm]
13 TeV& ATLAS-CONF-2017-039~\cite{ATLAS:2017uun} & $2\;$-$\;3\ell + \met$ & 36.1 \\[0.3mm]
\hline
\end{tabular}
\caption{List of the searches included in \checkmate which are most relevant  for our models. The first column refers to the 
center-of-mass energy, 8 or 13 TeV. The second column is the name of the analysis with the reference. The final state studied 
in each analysis is shown in the third column, while the fourth indicates the integrated luminosity used in each search, respectively. }
\label{tab:searches}
\end{table}

In order to determine if a point is excluded or not by a search, {\tt CheckMATE} compares the estimate of  the number of signal 
events with the 95\% C.L. observed limit 
\begin{eqnarray}
r= \frac{S-1.96\cdot\Delta S}{S^{95}_{\rm exp}},
\label{eq:rsig}
\end{eqnarray}
here $S$ stands for the number of signal events within \texttt{CheckMATE}, $\Delta S$ is the uncertainty due to MC 
errors\footnote{In our case we assume that the MC uncertainty to the signal events is only statistical, so it is given by $\Delta 
S=\sqrt{S}$.} and $S^{95}_{\rm exp}$ is the 95\% C.L. limit on signal events imposed by the experiment. A point is excluded 
if the value of $r$ is larger than 1.\footnote{We do not totally control all the aspects relevant for a true simulation, like systematic 
errors and higher order corrections. In order to take these uncertainties into account one can define a non-conclusive region 
defined as the area between $1.5>r>0.67$ where a point cannot be fully allowed or excluded.} $r$ is computed for every signal 
region of every analysis and then the best exclusion limit is chosen, taking the one that presents the best expected exclusion 
potential. This choice can result in  the total exclusion limit being weaker than the limit from a single 
search in one specific parameter area. Furthermore,  {\tt CheckMATE} does not combine searches or signal regions. Thus the 
limits such determined are conservative.


\section{Numerical Results}
\label{sec:numerics}

After having defined under which conditions we can consider a model point excluded, we apply these considerations to the 
different models.


\subsection{Model Ia}

\begin{figure}[t]
	\begin{center}
		\includegraphics[width=.85\linewidth]{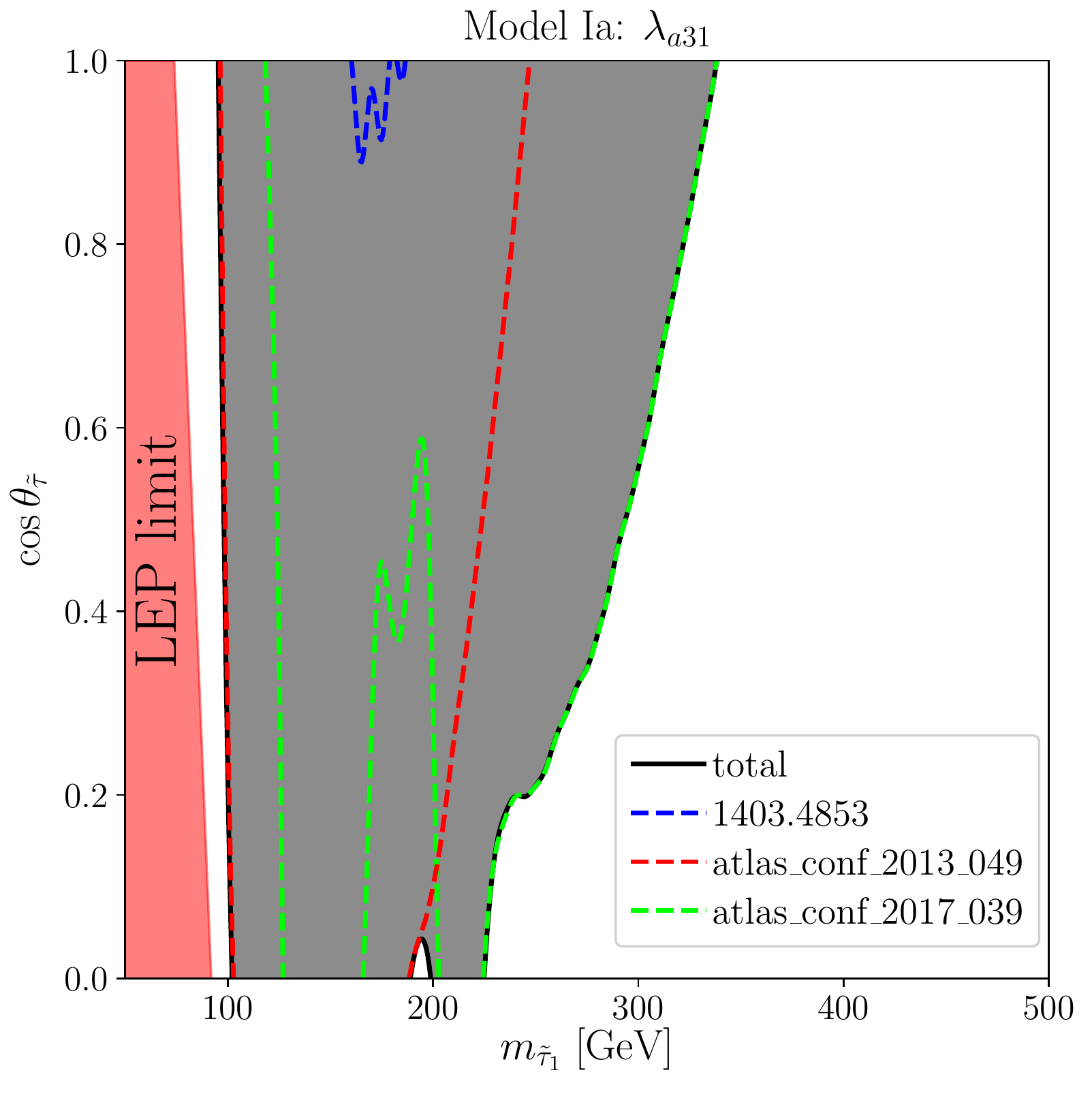}
	\end{center}
	\caption{Exclusion regions for Model Ia as a function of the stau mass and the mixing angle. The different 
	excluded regions due to the different analyses are shown as dashed lines of different colors. The full black line depicts the total exclusion region that is shown as a gray area. The LEP lower stau mass limit in RPV is shown as a red area on the left.}
	\label{fig:a31}
\end{figure}

In Model Ia, the relevant operator is $L_aL_3\bar{E}_1$, $a=1,2$. As we see in Tab.~\ref{tab:signatures},  at leading order, the 
stau decays exclusively as $\tilde{\tau}^\pm \to e^\pm \nu_a$. The final state signature is two opposite sign electrons plus missing 
energy. The results of testing this model against \texttt{CheckMATE} are depicted in Fig.~\ref{fig:a31}. The exclusion contours, 
requiring $r\geq1$, are shown in the ($m_{\tilde\tau_1},\cos\theta_{\tilde\tau}$) plane, as dashed colored curves. The different colors 
represent the different analyses. The full black line represents the total exclusion limit, which encompasses the 
gray area.

The blue-dashed line denotes the exclusion from direct stop production followed by leptonic decays at $\sqrt{s}=8\,$TeV of 
Ref.~\cite{Aad:2014qaa}.  This analysis excludes only a small parameter range around $m_{\tilde\tau_1}=150-
180\,$GeV and $\cos\theta_{\tilde\tau}>0.9$. It has only a weak sensitivity, as it was designed to look for same- and 
different-flavour final state leptons, while here only electrons are present. 

The red-dashed line corresponds to the two-lepton analysis at $\sqrt{s}=8\,$TeV of Ref.~\cite{TheATLAScollaboration:2013hha}. Overall
 this analysis is not designed for light staus below about 100\,GeV in mass due to the cuts implemented in the search. The power of this
search has a mild dependence on the mixing angle, at the upper mass end. This is because right-handed staus have a smaller 
production cross section than left-handed staus by about a factor of two in this mass range~\cite{Fuks:2013lya,Fiaschi:2018xdm}.  For
pure $\tilde\tau_R$, $\cos\theta_{\tilde \tau}=0 $, the exclusion of this search reaches up to $m_{\tilde\tau_1}>185\,$GeV. The lower 
mass bound increases as the stau becomes mixed, up to a mass of $240\,$GeV for a pure $\tilde\tau_L$.

The green-dashed line corresponds to the 2-3 leptons plus missing energy analysis at 13\,TeV and with an integrated luminosity of $
\mathcal{L}=36.1$fb$^{-1}$~\cite{ATLAS:2017uun}. This analysis has the highest stau mass sensitivity. For $\cos\theta_{\tilde\tau}
=0$ it reaches upto masses of 225\,GeV, and for $\cos\theta_{\tilde\tau}=1$ it extends all the way upto 322\,GeV. However, 
there is a gap in the sensitivity for stau masses between 165 and $200\,$GeV ranging upto $\cos\theta_{\tilde\tau}=0.6$. This is mainly 
because for  these points the observed events in the experiment are fewer than the expected ones, while in the other regions the 
observed number of events is bigger than the expected one. This range is mostly covered by the 
analysis~\cite{TheATLAScollaboration:2013hha}, the red-dashed curve. Furthermore, the analysis corresponding to the green-dashed 
curve also fails for stau masses below about 125\,GeV since the cuts are too strict to allow for sensitivity to lighter stau masses.

In red we show on the left the lower limit on the stau mass obtained at LEP, as discussed in Sec.~\ref{sec:LEP-Bounds}. There
is a significant gap to the LHC sensitivity.

The total exclusion limit, the combination of the excluded regions, is presented as a full black line, with the enclosed area in dark 
gray. For $\theta_{\tilde\tau}=0$, \textit{i.e.} for the lightest stau being pure $\tilde\tau_L$, we can exclude masses between about 
100\,GeV and 322\,GeV. At $\theta_{\tilde\tau}=\frac{\pi}{2}$ the upper range is reduced to about 225\,GeV, with a small search 
gap just below 200\,GeV in mass. There is a significant gap to the LEP bound at low mass, which is larger at
$\cos\theta_{\tilde\tau}=1$. The LHC sensitivity is higher at 
$\theta_{\tilde\tau}=0$, since the production cross section is higher for pure $\tilde\tau_L$.


\subsection{Model Ib}

\begin{figure}[ht]
	\begin{center}
		\includegraphics[width=.85\linewidth]{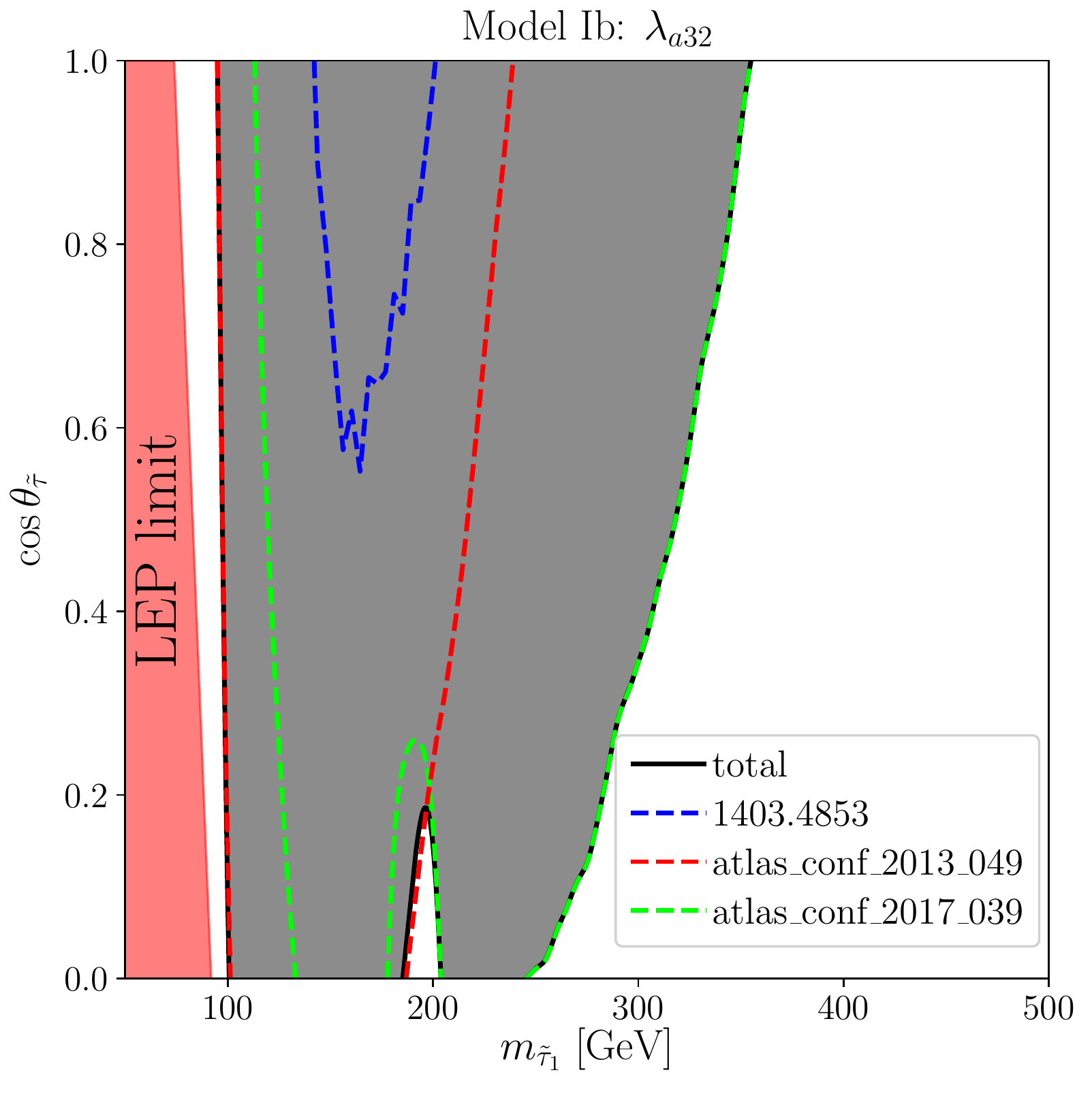}
	\end{center}
	\caption{Exclusion regions for Model Ib as a function of the stau mass and the mixing angle. The code of lines, areas and colors are the same as Fig.~\ref{fig:a31}.}
	\label{fig:a32}
\end{figure}

The operator that defines the Model~Ib is $L_aL_3\bar{E}_2$, $a=1,2$. The principal signature is two opposite sign muons plus missing
energy, \textit{cf.}  Tab.~\ref{tab:signatures}. In Fig.~\ref{fig:a32} we show the results of testing this model against {\tt CheckMATE}. The
same color and line code  is used as in Fig.~\ref{fig:a31}. Here the direct stop production search (blue-dashed line) is more
sensitive than in the previous case, due to the higher efficiency muon detection. However, this search is not competitive, compared to 
the other two. The red-dashed line, corresponding to the two-lepton search at 8 TeV, constrains stau masses in the range
$m_{\tilde{\tau}_1}=100-180\,,$GeV for a mixing angle $\cos\theta_{\tilde \tau}=0$ and $m_{\tilde{\tau}_1}=100-245\,$GeV for a 
mixing angle $\cos\theta_{\tilde \tau}=1$. The latter is more restrictive, since $\tilde\tau_L$ production
is larger than $\tilde\tau_R$.

The 2-3 lepton search at $\sqrt{s}=13$ TeV (green-dashed line) has a better reach in terms of mass exclusion. However, for staus 
that are mostly right handed, $\tilde{\tau}_1 \to \tilde{\tau}_R$, {\textit{i.e.}} $\cos\theta_{\tilde \tau}\to 0$, the same behaviour as in 
Model Ia arises, and a small gap just below $m_{\tilde{\tau}_1}=200\,$GeV appears. Half of this gap is covered by the 2 lepton 
search at 8 TeV (red-dashed line). As  in Model Ia, the 2-3 lepton search at 13\,TeV cannot cover well the low mass region due to 
stricter cuts in the search, and again the 2 lepton search can reach those lower values of the mass. The total exclusion line (solid 
black) is able to exclude stau masses $m_{\tilde{\tau}_1}=100-240\,$GeV for $\cos\theta_{\tilde\tau}=0$, modulo the small gap, 
and $m_{\tilde{\tau}_1}=100-345\,$GeV for $\cos\theta_{\tilde \tau}=1$. Again there is a significant gap between the lower LEP 
limit, \textit{cf.} Sec.~\ref{sec:LEP-Bounds}, and the low-mass exclusions from the LHC, for all mixing angles.


\subsection{Model II}

\begin{figure}[t!]
	\begin{center}
		\includegraphics[width=.85\linewidth]{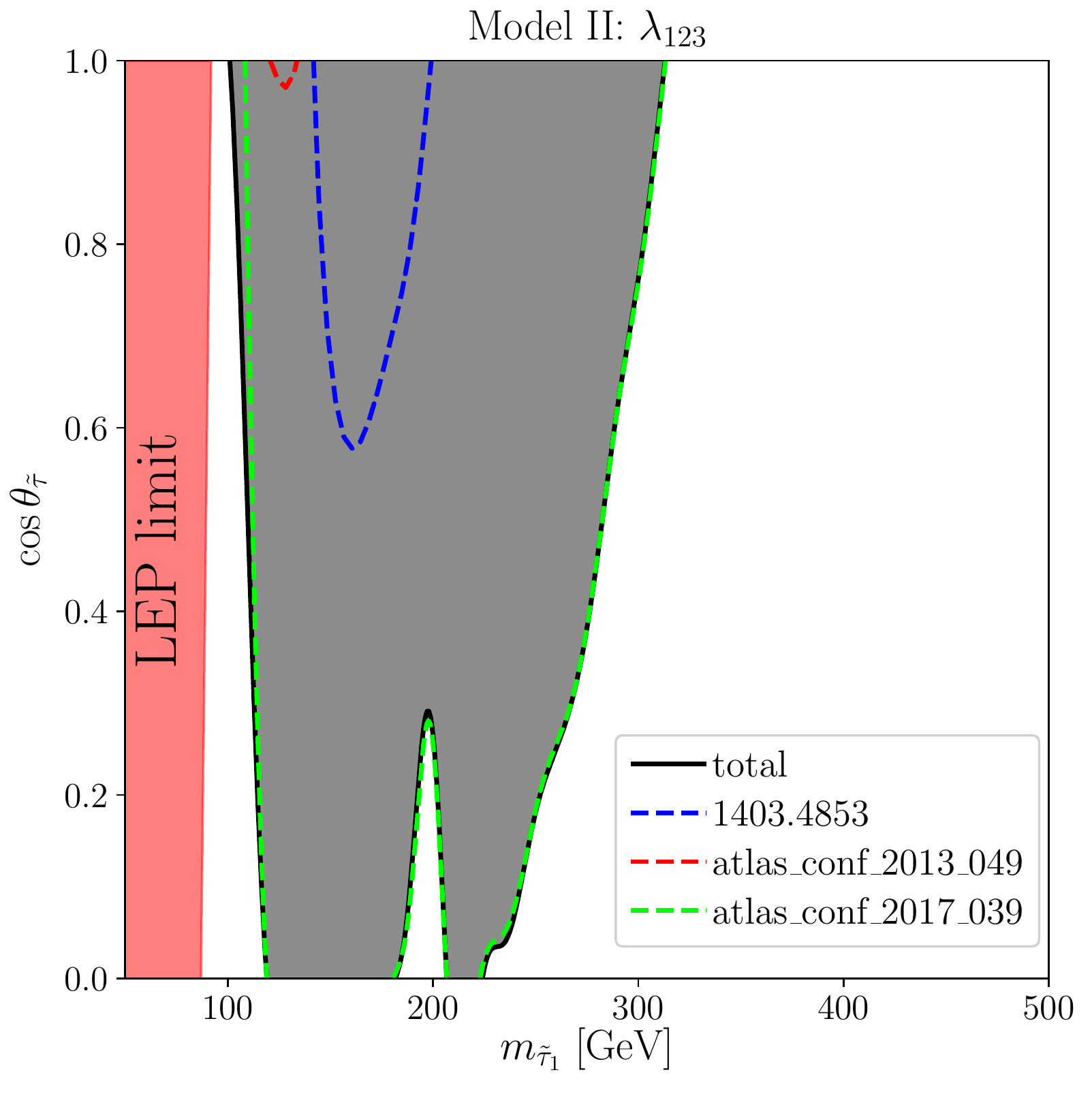}
	\end{center}
	\caption{Exclusion regions for Model II as a function of the stau mass and the mixing angle. The code of lines, areas and colors 
	are the same as Fig.~\ref{fig:a31}.}
	\label{fig:123}
\end{figure}

The relevant operator for  Model II is $L_1L_2\bar{E}_3$. The stau decay is either $\tilde{\tau}_1^+\to e^+\bar{\nu}_ \mu$ or $\tilde{\tau}
_1^+\to \mu^+ \bar{\nu}_e$, and the final state signatures of stau pair production are $(e^+e^-/ \mu^+\mu^-/ e^\pm\mu^\mp) + \met$. In
Fig.~\ref{fig:123} the results of testing this model against LHC searches in \texttt{CheckMATE} are shown. The stop pair production 
search (blue) is now more effective than for Models Ia, Ib. This is due to the inclusion of different flavour lepton signatures. However, it 
is still not relevant for the final exclusion region. The two lepton search at 8\,TeV (red) drops drastically in sensitivity, in comparison with 
both cases of Model I. One reason is that the different flavours appearing in the final state lower the detectability in this specific search. 
In this case, as in the previous ones the most powerful in terms of constraining power is the 2-3 lepton search at 13\,TeV (green). This 
search is also co\"incident with the total exclusion line (black) for this model. Almost pure left-handed staus, $\cos\theta_{\tilde \tau}\to 1$,
are excluded for the mass range $m_{\tilde{\tau}_1}=100-305\,$GeV, while in the case of right-handed staus, $\cos\theta_{\tilde \tau}\to 
0$, the excluded mass range is $m_{\tilde{\tau}_1}=120-180\,$GeV and $m_{\tilde{\tau}_1}= 205 -225\,$GeV. There is also a mass gap 
in this model between $m_{\tilde{\tau}_1}=180-205\,$GeV. In this case the gap is not partially covered by other searches. Again there is 
a significant gap at low stau masses above the LEP bound,  \textit{cf.} Sec.~\ref{sec:LEP-Bounds}.


\subsection{Model IIIa}

\begin{figure}[ht]
	\begin{center}
		\includegraphics[width=.85\linewidth]{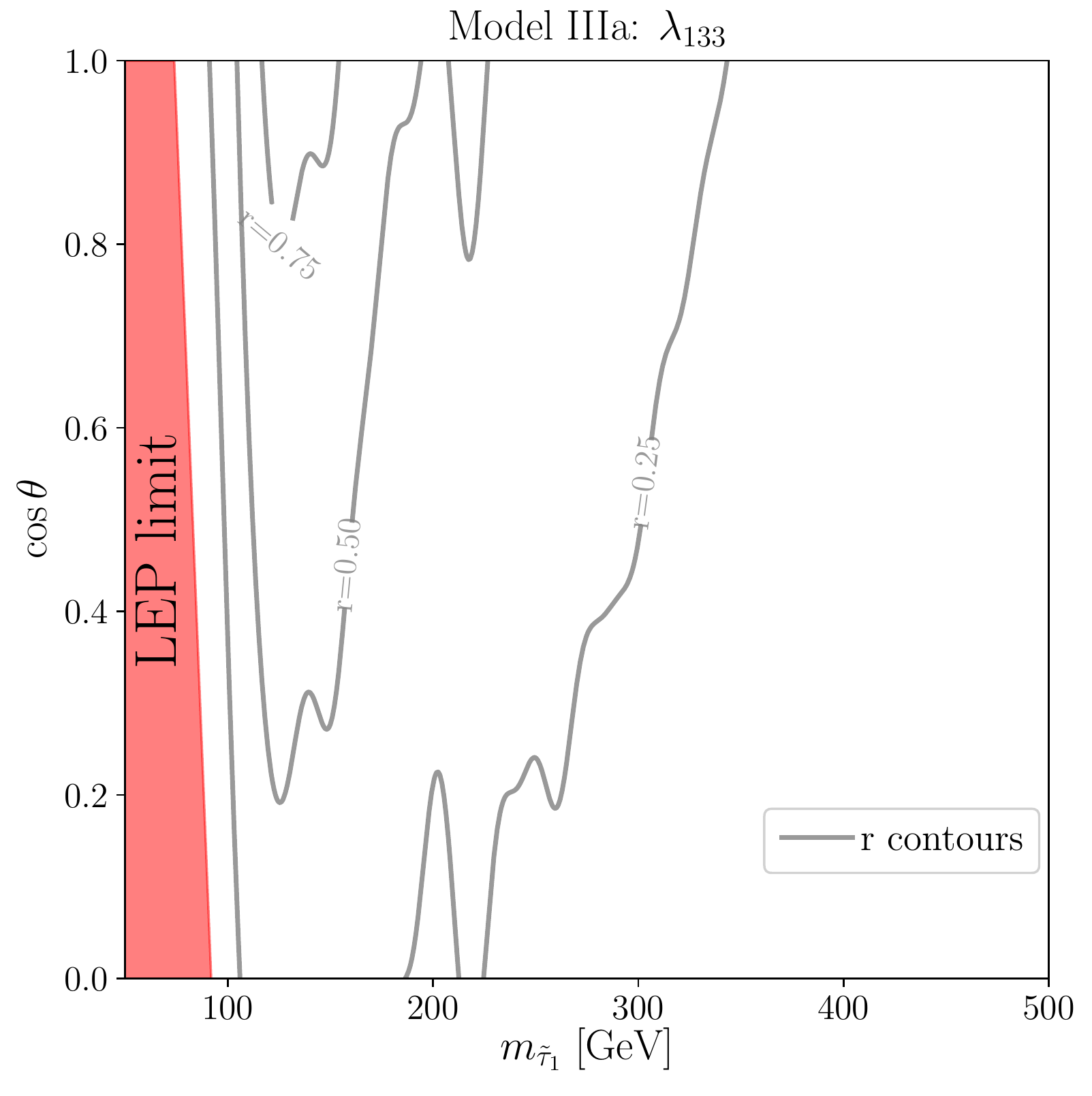}
	\end{center}
	\caption{Recasting of Model IIIa using \texttt{CheckMATE}. We show iso-contours of $r$, \textit{cf.} Eq.~(\ref{eq:rsig}),
	for $r=0.25,0.50$ and $0.75$, as a function of the stau mass and the mixing angle. }
	\label{fig:133}
\end{figure}

Model IIIa corresponds to the dominant operator $L_1L_3\bar{E}_3$. The stau decays as $\tilde{\tau}_1^+\to (\tau^+\nu_e,\, 
\tau^+\bar{\nu}_e,\, e^+\nu_\tau)$, \textit{cf.} Tab.~\ref{tab:signatures}. The branching ratios are given in Eqs.~(\ref{eq:stau-Br1}), and
(\ref{eq:stau-Br2}), as a function of the mixing angle. In Models I and II the staus decay 100\% to charged electrons or muons. In
Models IIIa and below in IIIb at least 50\% of the two-body decays are to $\tau$'s, depending on the mixing angle. This significantly 
degrades the experimental sensitivity, since the implemented searches do not involve $\tau$-signatures. In Fig.~\ref{fig:133} we show
the results of recasting this model in \texttt{CheckMATE}, in particular iso-contours of $r$, \textit{cf.} Eq.~(\ref{eq:rsig}). We do 
not find any region with $r\geq1$ and therefore the current searches are not sensitive enough to constrain this model. The most 
sensitive area, $r>0.5$ in the $(m_{\tilde{\tau}_1}, \cos\theta_{\tilde \tau})$ plane is roughly for $100\,\mathrm{GeV}<m_{\tilde{\tau}_1}
<200\,$GeV and $\cos\theta_{\tilde\tau}>0.25$. This results mainly from the ATLAS-CONF-2017-039~\cite{ATLAS:2017uun} search. 
This leads us to think that  in the near 
future this model will be tested with a similar sensitivity to the other models. Currently the LEP limits are the strictest on this
model,  \textit{cf.} Sec.~\ref{sec:LEP-Bounds}.

\begin{figure}[h!]
	\begin{center}
        \includegraphics[width=.85\linewidth]{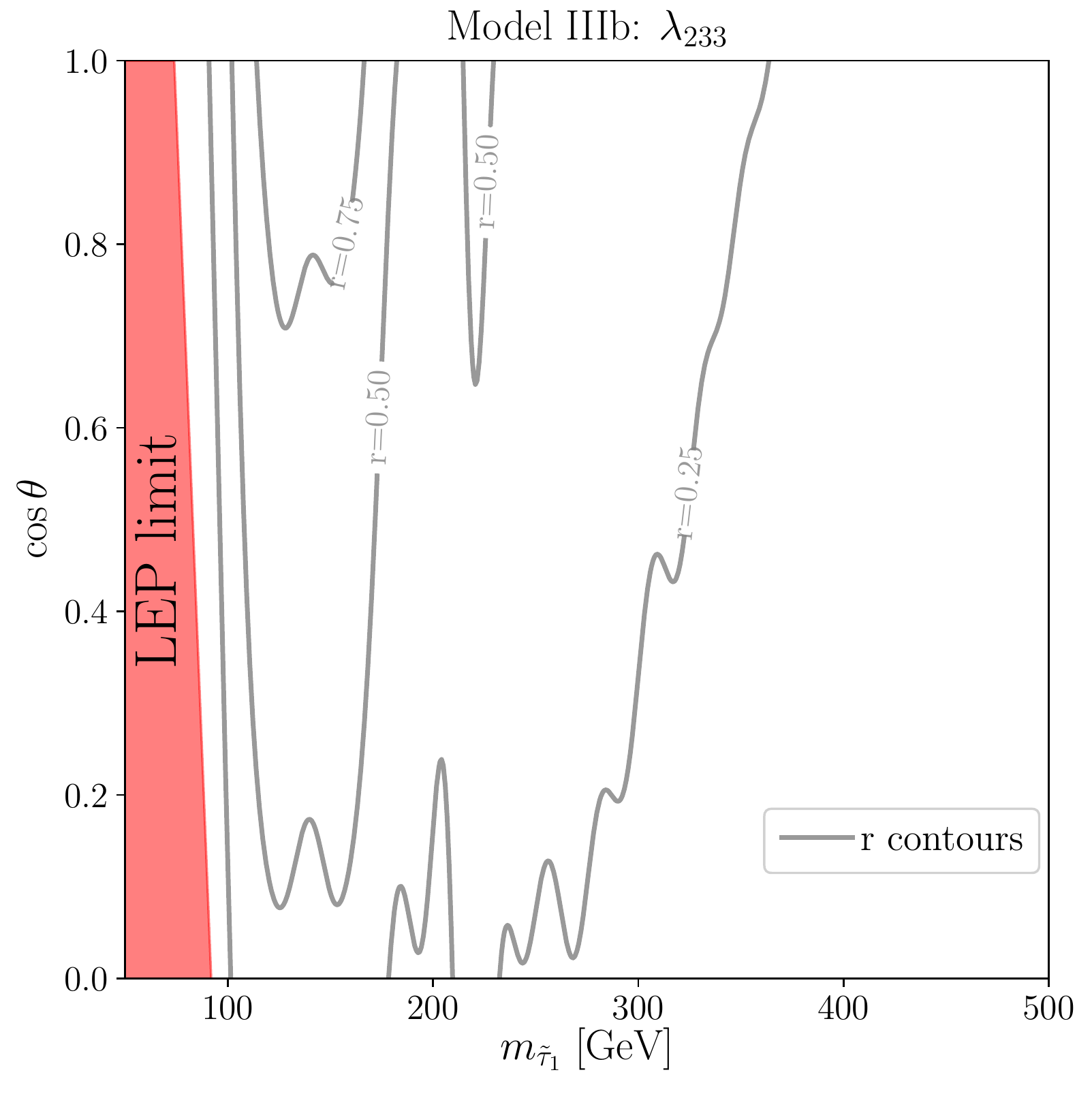}
	\end{center}
\caption{As in Fig.~\ref{fig:133} but for Model IIIb. }
\label{fig:233}
\end{figure}\subsection{Model IIIb}

The relevant operator for Model IIIb is $L_2L_3\bar{E}_3$. This case is  similar to Model IIIa, with the electrons in the final 
state of the stau decay replaced by muons, \textit{cf.} Tab.~\ref{tab:signatures}. The maximum branching ratio to muons is also 50\%, 
as for the electrons in Model IIIa. The efficiency for muons is higher than for electrons and we expect a higher sensitivity. However,
also in this case the searches implemented in \texttt{CheckMATE} are not sensitive enough to constrain Model IIIb. For that reason 
we depict in Fig.~\ref{fig:233} iso-contours for $r=0.25,0.50$ and $0.75$. The more promising region, $r\geq0.5$, is as expected
slightly larger than in Model IIIa, reaching lower values of $\cos\theta_{\tilde \tau}\geq0.1$. We expect that future searches will be 
sensitive to this scenario. Again, currently the LEP limits are the strictest on this
model,  \textit{cf.} Sec.~\ref{sec:LEP-Bounds}.


\section{Conclusions}
\label{sec:conclusions}
We have examined the the pair production of supersymmetric staus followed by the direct RPV decay via $L_iL_j\bar E_k$ operators at 
the LHC. We assume these decays comprise 100\% of the branching ratios, corresponding to assuming the stau is the LSP, which is 
indeed the case for a large range of parameters in the RPV-CMSSM. To compare with data we have employed the program \texttt{ 
CheckMATE}. We have demonstrated that current data from LHC set significant bounds on the lightest stau mass and the relevant stau
mixing angle. The stau LSP decays via the $LL\bar{E}$ into SM leptons. Therefore existing experimental searches for new physics in
multileptonic channels can significantly constrain such scenarios. Current searches are able to put lower limits on the mass of 
the staus, with however a significant gap between the lower LEP limit and the onset of current LHC sensitivity. For the scenario where 
the stau can only decay into electrons and neutrinos ($L_{1,2}L_3\bar E_1$) the mass exclusion limit is set to $m_{\tilde{\tau}_1}> 225$ 
(322)\,GeV for right-handed (left-handed) staus. If the staus decay only into muons and neutrinos 
($L_{1,2}L_3\bar E_2$), the limits are $m_{\tilde{\tau}_1}> 240$ (345)\,GeV for right-handed (left-handed) staus. When the decay into 
both electrons and muons is open ($L_1L_2\bar E_3$), then the searches are less efficient. In the case of pure left-handed staus the 
mass limit is $m_{\tilde{\tau}_1}> 305\,$GeV. However, in the case of pure right-handed staus the lower mass limit
is $m_{\tilde{\tau}_1}> 225\,$GeV, with however a significant gap in sensitivity between 180\,GeV and 205\,GeV. And as in all cases there
is a gap between the lower LEP limit and the onset of the LHC sensitivity. For a detailed stau mixing angle dependence of all above
bounds see Figs.~\ref{fig:a31}-\ref{fig:123}.

In the scenarios where the stau decays to tau leptons with at least 50\% branching ratio ($L_{1,2}L_3\bar E_3$), the current searches 
implemented in \texttt{CheckMATE} are not sensitive enough to set limits on the mass of the stau. We expect that in future 
runs of the LHC most of the parameter space could be explored by new multilepton searches.

\section*{Acknowledgements}

We would like to thank Manuel Krauss for interesting discussions and collaboration in the initial phase of this project. HKD  and VML 
acknowledge support of the BMBF-Verbundforschungsprojekt 05H18PDCA1. HKD thanks SCIPP at UCSC for kind hospitality, while part of this work was completed. VML acknowledges support by the Deutsche Forschungsgemeinschaft (DFG, German Research Foundation) under Germany's Excellence Strategy - EXC 2121 ``Quantum Universe'' - 390833306.

\bibliographystyle{apsrev4-1}
\bibliography{lit}

\end{document}